\documentclass[twocolumn,prd,aps,superscriptaddress,nofootinbib,floatfix]{revtex4-2}

\usepackage{amsmath,amssymb,amsfonts}
\usepackage{bm}
\usepackage{braket}
\usepackage{graphicx}
\usepackage{xcolor}
\usepackage[colorlinks=true,linkcolor=blue,citecolor=blue,urlcolor=blue]{hyperref}

\begin{document}

\title{Magic Is in the Eye of the Measure}

\author{Zhi-Heng Shen}
\affiliation{School of Physical Sciences, University of Science and Technology of China, Hefei 230026, China}
\author{Zhewei Yin}
\affiliation{Institute of High Energy Physics, Chinese Academy of Sciences, Beijing 100049, China}
\author{Hao-Lin Zhang}
\affiliation{School of Physical Sciences, University of Science and Technology of China, Hefei 230026, China}

\begin{abstract}
Magic, or nonstabilizerness, is the quantum resource behind universal
quantum computation. Its applications in high-energy and nuclear physics
rely almost exclusively on the stabilizer R\'enyi entropy (SRE), although
magic has no unique measure. For two qubits, we compare the SRE with
three measures defined through optimizations over stabilizer states: the
min- and max-relative entropies of magic and the logarithmic free
robustness. We determine the maximal-magic states of all four measures,
with exact proofs, and find that they depend on the measure. In
the determination of the weak mixing angle from minimal magic in
M{\o}ller scattering, the min-relative entropy gives an
energy-dependent value that differs from that of the SRE, whereas the
max-relative entropy and the logarithmic free robustness give
$\sin^2\theta_W=1/4$ at all energies. We trace this difference to the
non-analytic structure of the optimization-based measures. Physical
conclusions drawn from extremizing magic can thus depend on the measure.
\end{abstract}

\maketitle

\section{Introduction}
\label{sec:intro}

Entanglement alone does not make a quantum computer powerful. By the
Gottesman--Knill theorem, circuits built from Clifford gates acting on
stabilizer states can be simulated efficiently on a classical computer,
however much entanglement they generate~\cite{Gottesman:1998hu,Aaronson:2004xuh}. Universal quantum computation requires an
additional resource, known as magic or nonstabilizerness, which is
supplied by non-Clifford operations or by magic states~\cite{Bravyi:2004isx,Gottesman:1999tea}. Measures of magic quantify
nonstabilizerness and are related to the cost of specific classical
simulation methods and resource-conversion tasks. Their study has
developed into a resource theory with a variety of
measures~\cite{Veitch:2012ttw,Emerson:2013zse,Howard:2017maw,Bravyi:2018ugg,Seddon:2021rij,Liu:2020yso,Leone:2021rzd,Chitambar:2018rnj}.

Over the past two years, magic has also emerged as a new lens on
high-energy and nuclear physics. It has been studied in top-quark pair
production at colliders~\cite{White:2024nuc,Aoude:2025jzc,Afik:2026pxv}, where it
has recently been observed by the CMS
Collaboration~\cite{CMS:2025cim}, in nuclear and hypernuclear
forces~\cite{Robin:2024oqc,Robin:2025dgw,Low:2026evp,Li:2026udy},
the structure of nuclei~\cite{Brokemeier:2024lhq}, neutrino
physics~\cite{Chernyshev:2024pqy}, QED and electroweak
scattering~\cite{Liu:2025qfl,Li:2026zum,Alvarado:2026owd}, gluon and
graviton scattering~\cite{Gargalionis:2025iqs,Gargalionis:2026onv,Chu:2026yxm,Nunez:2025xds},
extended Higgs sectors~\cite{Busoni:2025dns}, Higgs
decays~\cite{Viaux:2026skf}, deep-inelastic
scattering~\cite{Cheng:2025zaw}, lattice gauge
theories~\cite{Tarabunga:2023ggd,Falcao:2024msg,Ebner:2025pdm}, and quantum field
theory~\cite{Nystrom:2024oeq,Cepollaro:2024sod}; see
Ref.~\cite{Robin:2026lqp} for a recent review.
Particularly intriguing is the possibility that magic could single out
the parameters of the Standard Model, in the spirit of earlier proposals
based on entanglement~\cite{Cervera-Lierta:2017tdt,Beane:2018oxh,
Quinta:2022sgq,Thaler:2024anb,Liu:2025iwh,Cao:2026aye}. It was found
that the measured weak mixing angle tends to minimize the magic generated
in M{\o}ller scattering~\cite{Liu:2025bgw}. In
tree-level quark--antiquark scattering, a finite-coupling minimum of magic
production in the color-singlet top-quark channel was found to track the
running strong coupling, with the other inputs held
fixed~\cite{Liu:2026gju}.

Almost all of these developments use a single measure of magic, the
stabilizer R\'enyi entropy (SRE)~\cite{Leone:2021rzd}, or quantities derived
from it; the few exceptions employ mana for qutrit systems or for
detectors coupled to quantum fields~\cite{Durupt:2025wuk,Nystrom:2024oeq}. The
SRE has a clear practical advantage: it is defined in terms
of Pauli expectation values without any optimization over stabilizer
states.
However, unlike the entanglement of pure bipartite states, which is
uniquely quantified by the entanglement entropy in the asymptotic
limit, magic has no single preferred measure. Different measures are tied to different
operational tasks, such as the overlap with stabilizer states, the cost
of stabilizer decompositions in classical simulation, the sampling
overhead of quasiprobability methods, and stabilizer property
testing~\cite{Howard:2017maw,Bravyi:2018ugg,Liu:2020yso,Bittel:2025yhq}. It is
therefore natural to ask whether physical conclusions drawn from the SRE
also hold for other measures.

In this work, we address this question for two qubits,
the setting of most high-energy physics (HEP)
applications. We compare the second SRE with three measures defined
through optimizations over the stabilizer polytope: the min- and
max-relative entropies of magic, $\mathfrak{D}_{\min}$ and
$\mathfrak{D}_{\max}$~\cite{Datta:2009htm,Datta:2009yxd,Liu:2020yso,Bravyi:2018ugg,Regula:2017mvm},
and the logarithmic free robustness,
LR~\cite{Vidal:1998ch,Howard:2017maw,Liu:2020yso}. While all four
measures agree on the maximal-magic states of a single qubit, they do not
for two qubits, where the maximizers of LR even form a continuous family.
For all four measures, we determine the maximal value and the complete
set of maximizers and prove them exactly, including the maxima of the
SRE and of LR, which were previously known only
numerically~\cite{Liu:2025frx,Howard:2017maw}. As a physics application,
we revisit the determination of the weak mixing angle from minimal
magic~\cite{Liu:2025bgw}. The result depends qualitatively on the
measure, a difference we trace back to the non-analytic structure of the
measures.

\section{Magic and its measures}
\label{sec:measures}

For $n$ qubits, the Pauli group $\mathcal{P}_n$ is generated by tensor
products of the $2\times2$ identity $I$ and the Pauli matrices $X$, $Y$,
and $Z$. The Clifford group $\mathcal{C}_n$ is the normalizer of
$\mathcal{P}_n$ in the unitary group. It is generated by the Hadamard
gate $H$, the phase gate $S$, and the controlled-NOT gate
$\mathrm{CNOT}$, whose explicit forms are given in
App.~\ref{app:domain}. Pure stabilizer states, $\mathrm{STAB}_n$, are the
states $C\ket{0}^{\otimes n}$ with $C\in\mathcal{C}_n$; there are 6
of them for $n=1$ and 60 for
$n=2$~\cite{Garcia:2017uqv,Robin:2024oqc,Liu:2025qfl}. The
free states of the resource theory are the convex hull $\mathcal{F}$ of
$\mathrm{STAB}_n$, the stabilizer polytope, and the free operations
include Clifford unitaries, Pauli measurements, and classical
randomness~\cite{Emerson:2013zse,Chitambar:2018rnj}. A magic measure
$\mathcal{M}$ is a function that vanishes on $\mathcal{F}$, is invariant
under Clifford unitaries, and does not increase under free operations.

The first measure we consider is the \textbf{stabilizer R\'enyi entropy}
(SRE)~\cite{Leone:2021rzd}. For a pure state $\ket{\psi}$ of $n$ qubits, the
second SRE is
\begin{equation}
  M_2(\psi) = -\log_2 \frac{1}{2^n}\sum_{P\in\mathcal{P}_n/\{\pm1,\pm i\}}
  \bra{\psi}P\ket{\psi}^4 ,
  \label{eq:M2}
\end{equation}
where the sum runs over the $4^n$ Pauli strings, i.e., the elements of
$\mathcal{P}_n$ up to phases.\footnote{We use logarithms to base 2
throughout, so that all measures are given in bits. Some works use the
natural logarithm instead, e.g., Ref.~\cite{Liu:2025frx}; their values
are smaller than ours by a factor $\ln2$.} $M_2$ is a monotone for pure
states~\cite{Leone:2024lfr}.

The second and third measures are the \textbf{min-relative entropy of
magic} and the \textbf{max-relative entropy of magic}, the magic
counterparts of the min- and max-relative
entropies~\cite{Datta:2009htm,Datta:2009yxd}, in the notation of
Ref.~\cite{Liu:2020yso},
\begin{align}
  \mathfrak{D}_{\min}(\rho) &= -\log_2 \max_{\sigma\in\mathcal{F}}
     \mathrm{Tr}\left(\Pi_\rho\,\sigma\right), \\
  \mathfrak{D}_{\max}(\rho) &= \log_2 \min\left\{\lambda :
     \rho\le\lambda\sigma,\ \sigma\in\mathcal{F}\right\},
\end{align}
where $\Pi_\rho$ is the projector onto the support of $\rho$. For pure
states they reduce to~\cite{Liu:2020yso,Regula:2017mvm}
\begin{align}
  \mathfrak{D}_{\min}(\psi) &= -\log_2 F_{\mathrm{STAB}}(\psi), \label{eq:Dmin}\\
  \mathfrak{D}_{\max}(\psi) &= \log_2 \xi(\psi), \label{eq:Dmax}
\end{align}
where
\begin{equation}
  F_{\mathrm{STAB}}(\psi)=\max_{s\in\mathrm{STAB}_n}
     |\braket{s|\psi}|^2
\end{equation}
is the stabilizer fidelity~\cite{Bravyi:2018ugg}, i.e., the largest
overlap with a stabilizer state $\ket{s}$, and
\begin{equation}
  \xi(\psi)=\min\Big\{\Big(\sum_s|a_s|\Big)^2 :
     \ket{\psi}=\sum_{s\in\mathrm{STAB}_n}a_s\ket{s}\Big\}
\end{equation}
is the stabilizer extent~\cite{Bravyi:2018ugg}, the minimal cost of a
decomposition of $\ket{\psi}$ into stabilizer states.

The fourth measure is the \textbf{logarithmic free robustness},
$\mathrm{LR}(\rho)=\log_2[1+R(\rho)]$, where
\begin{equation}
  R(\rho)=\min\left\{\nu\ge0:\rho=(1+\nu)\sigma_1-\nu\sigma_2,\
  \sigma_{1,2}\in\mathcal{F}\right\}
  \label{eq:R}
\end{equation}
is the free robustness of magic~\cite{Liu:2020yso}, the magic counterpart
of the robustness of entanglement~\cite{Vidal:1998ch}. It is related to the
robustness of magic defined in Ref. \cite{Howard:2017maw},
$\mathcal{R}(\rho)=\min\{\sum_s|q_s|:\rho=\sum_s q_s\ket{s}\!\bra{s},\
q_s\in\mathbb{R}\}$, by
$\mathcal{R}=1+2R$.

The three optimization-based measures obey
$\mathfrak{D}_{\min}\le\mathfrak{D}_{\max}\le\mathrm{LR}$.
All three are
quasi-convex,
$\mathcal{M}(\sum_ip_i\rho_i)\le\max_i\mathcal{M}(\rho_i)$, so their
maxima over all states are attained on pure
states~\cite{Liu:2020yso,Howard:2017maw}. In what follows we
therefore restrict ourselves to pure states, for which all four measures
are defined.

\begin{figure}[t]
  \centering
  \includegraphics[width=\columnwidth]{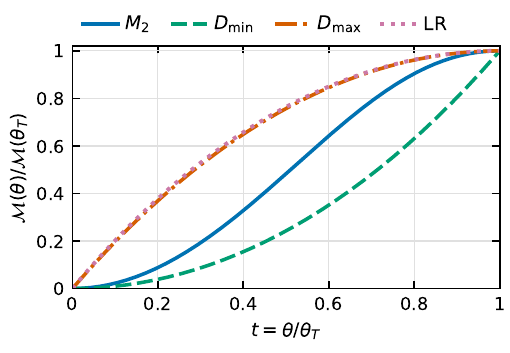}
  \caption{Four magic measures along the single-qubit path from the
  stabilizer state $\ket{0}$ ($t=0$) to the $T$-type state ($t=1$),
  $\varphi=\pi/4$, $t=\theta/\theta_T$. Each curve is normalized to its
  value at the $T$-type state.}
  \label{fig:1qubit}
\end{figure}

The differences among the four measures are already visible for a single
qubit,
$\ket{\psi}=\cos\frac{\theta}{2}\ket{0}+e^{i\varphi}\sin\frac{\theta}{2}\ket{1}$,
for which all four are given in closed form in App.~\ref{app:1qubit}.
All four are maximized by the $T$-type states~\cite{Bravyi:2004isx},
whose Bloch vectors $\bm{r}=(x,y,z)$ have $|x|=|y|=|z|=1/\sqrt{3}$.
Figure~\ref{fig:1qubit} follows the path $\varphi=\pi/4$,
$0\le\theta\le\theta_T=\arctan\sqrt{2}$, from the stabilizer state
$\ket{0}$ to a $T$-type state.

Figure~\ref{fig:1qubit} reflects a basic difference between $M_2$ and
the other three measures. By Eq.~\eqref{eq:M2}, $2^{-M_2}$ is a
polynomial in the Pauli expectation values, so $M_2$ is smooth
everywhere. The optimization-based measures, in contrast, may be
non-analytic wherever the optimal stabilizer state or decomposition
changes, including at their minima and maxima. At the stabilizer state
$\ket{0}$, $\mathfrak{D}_{\max}$ and LR grow linearly, $\propto\theta$,
inheriting the conical non-analyticity of the $\ell_1$ norm in their
stabilizer decompositions, whereas
$\mathfrak{D}_{\min}$, dominated by a single stabilizer overlap, grows
quadratically, like $M_2$. At the $T$-type state the roles are exchanged:
$\mathfrak{D}_{\min}$, a maximum over finitely many overlaps, reaches its
maximum with a finite slope, i.e., at a cusp, whereas the other three
measures approach it with vanishing slope. Both features have two-qubit
counterparts: cusp-like maxima of $\mathfrak{D}_{\min}$
(Sec.~\ref{sec:maximal}), and a cusp in the magic produced in M{\o}ller
scattering as a function of the weak mixing angle (Sec.~\ref{sec:ew}).

\section{Maximal magic of two qubits}
\label{sec:maximal}

We now determine the maximal-magic states of two qubits for the four
measures. The strategy is the same for all of them. A direct numerical
search first identifies the maximal value and the candidate maximizers,
and an exact certificate then proves that this value is the global
maximum and that the candidates are the only maximizers. The search
exploits Clifford invariance: every two-qubit state is Clifford-equivalent
to a state in a domain covering $1/768$ of the parameter space, which we detail in App.~\ref{app:domain};
the maximizers
found are then mapped back to the full state space with the Clifford
group. The exact certificates, given in
Apps.~\ref{app:M2proof}--\ref{app:LRproof}, are tailored to the
structure of each measure. $2^{-M_2}$ is a polynomial, which admits a
sum-of-squares certificate. $F_{\mathrm{STAB}}$ is the maximum of
finitely many quadratic functions, with cusp-like extrema, which we
handle with an exact local analysis combined with a branch-and-bound
search. The robustness is the value of a linear program, whose dual
reduces the problem to finitely many vertices.

For $M_2$, the maximal value is $M_2^{\max}=\log_2(16/7)\approx1.193$,
attained on 480 states. Their concurrence, a standard measure of
entanglement~\cite{Hill:1997pfa,Wootters:1997id}, which for
$\ket{\psi}=c_1\ket{00}+c_2\ket{01}+c_3\ket{10}+c_4\ket{11}$ reads
$C(\psi)=2|c_1c_4-c_2c_3|$, takes only the two values $1/\sqrt2$ and
$1/2$. Ref.~\cite{Liu:2025frx} provided strong numerical evidence
for this maximum (see also Ref.~\cite{Roman:2026mcy}), and we prove in
App.~\ref{app:M2proof} that it is the global maximum and that these 480
states are the only maximizers.

For $\mathfrak{D}_{\min}$ and $\mathfrak{D}_{\max}$, our search finds
that both measures reach their maximal value
\begin{equation}
  \mathfrak{D}_{\min}^{\max}=\mathfrak{D}_{\max}^{\max}
  =\log_2\frac{6}{2+\sqrt{3}}\approx0.6850
\end{equation}
on the \emph{same} set of 640 states, which form a single Clifford orbit
with representative
\begin{equation}
  \ket{\psi_{\mathfrak{D}}}=\frac{1}{\sqrt{3}}
  \left(e^{i\pi/6}\ket{01}+e^{i\pi/3}\ket{10}+\ket{11}\right).
  \label{eq:psiD}
\end{equation}
Of the 640 states, 64 are the products $\ket{T}\otimes\ket{T}$ of
single-qubit $T$-type states, while the other 576 are entangled with
concurrence $2/3$. These states belong to the class of
Clifford-symmetric states studied in Ref.~\cite{Erew:2025egp}, which
were identified as extremal points of several magic measures. The
coincidence of the two maxima and of their maximizers is not accidental.
For pure states, $F_{\mathrm{STAB}}^{1/2}$ and $\xi^{1/2}$ are a pair of
dual norms, so the minimum of one over the unit sphere is the inverse of
the maximum of the other, with identical
extremizers~\cite{Regula:2017mvm,Regula:2020ook}. It therefore suffices
to certify the minimum of $F_{\mathrm{STAB}}$. In App.~\ref{app:Dproof}
we prove that $F_{\mathrm{STAB}}\ge(2+\sqrt3)/6$ for every two-qubit pure
state, with equality only on these 640 states, each of which is a
cusp-like maximum of $\mathfrak{D}_{\min}$.

For LR, the maximal value
$\mathrm{LR}^{\max}=\log_2\frac{1+\sqrt5}{2}\approx0.6942$, i.e., a
robustness of magic $\mathcal{R}=\sqrt5$, was found numerically in
Ref.~\cite{Howard:2017maw} for a state lying maximally outside one of
the facets of the stabilizer polytope. Our search shows that, unlike for
the other measures, the maximum is attained on a continuum of states: 192
two-dimensional subspaces, or planes, of the two-qubit Hilbert space. One
such plane is spanned by the orthonormal vectors
\begin{equation}
\begin{aligned}
  \ket{e_1}&=\mathcal{N}\left(\tfrac{1-i}{1+\sqrt5},\
    -\tfrac{1-i}{1+\sqrt5},\ 1,\ 0\right)^{T},\\
  \ket{e_2}&=\mathcal{N}\left(\tfrac{1+i}{1+\sqrt5},\
    \tfrac{1+i}{1+\sqrt5},\ 0,\ 1\right)^{T},
\end{aligned}
  \label{eq:LRplane}
\end{equation}
with $\mathcal{N}=\sqrt{(5+\sqrt5)/10}$. The pure states in a plane form
a Bloch sphere, and all of them have $\mathrm{LR}=\mathrm{LR}^{\max}$.
Each plane contains exactly two product states, and the concurrence varies
between 0 and $\sqrt{3/5}$ within a plane. We prove the above in
App.~\ref{app:LRproof}, including mixed states, which are maximizers
exactly when they are supported on one of the planes. Further properties
of the planes are given in App.~\ref{app:LR}.

\begin{table}[t]
  \caption{Magic of the maximal-magic states of one measure (rows),
  evaluated with the other measures (columns), normalized to the maximal
  value of the column measure. Intervals indicate the range over an
  LR-maximizing plane.}
  \label{tab:cross}
  \begin{ruledtabular}
  \begin{tabular}{lcccc}
    Maximizers & $\mathfrak{D}_{\min}$ & $\mathfrak{D}_{\max}$ & LR & $M_2$ \\
    \hline
    $\mathfrak{D}_{\min,\max}$ & 1 & 1 & 0.997 & 0.981 \\
    LR & $[0.68,0.84]$ & $[0.84,0.96]$ & 1 & 0.888 \\
    $M_2$ & 0.990 & 0.990 & 0.977 & 1 \\
  \end{tabular}
  \end{ruledtabular}
\end{table}

The three sets of maximal-magic states, those of
$\{\mathfrak{D}_{\min},\mathfrak{D}_{\max}\}$, of LR, and of $M_2$, are
mutually disjoint. In contrast to the single-qubit case, ``maximal
magic'' for two qubits is therefore a measure-dependent notion.
Table~\ref{tab:cross} shows how much magic a maximizer of one measure
carries according to the others. The maximizers of $\mathfrak{D}_{\min}$
and $M_2$ are close to maximal in every measure; at the $M_2$ maximizers,
all three optimization-based measures take the common value
$\log_2(8/5)$, and at the $\mathfrak{D}$ maximizers $M_2=\log_2(9/4)$
and $\mathrm{LR}=\log_2[(3+2\sqrt3)/4]$.
The LR maximizers are the outliers: $M_2$ is constant on every
LR-maximizing plane, equal to $\log_2(25/12)$, and $\mathfrak{D}_{\min}$
varies strongly within a plane. A state that is maximally magic as
measured by LR can thus be rather unremarkable as measured by $M_2$ or
$\mathfrak{D}_{\min}$.

\begin{table}[t]
  \caption{Fubini--Study distances $d$
  between the maximal-magic sets $S_{\mathfrak{D}}$,
  $S_{\mathrm{LR}}$, and $S_{M_2}$. ``Min'' is the smallest distance
  between the two sets; ``Max'' is the largest distance from a state in
  $S_{\mathrm{LR}}$ to the nearest state in the other set.}
  \label{tab:dist}
  \begin{ruledtabular}
  \begin{tabular}{lcc}
    Sets & Min & Max\\
    \hline
    $S_{\mathfrak{D}}$, $S_{M_2}$ & $\pi/12$ & --- \\
    $S_{\mathfrak{D}}$, $S_{\mathrm{LR}}$ & 0.135 & 0.352\\
    $S_{\mathrm{LR}}$, $S_{M_2}$ & 0.232 & 0.359\\
  \end{tabular}
  \end{ruledtabular}
\end{table}

\begin{figure*}[t]
  \centering
  \includegraphics[width=\textwidth]{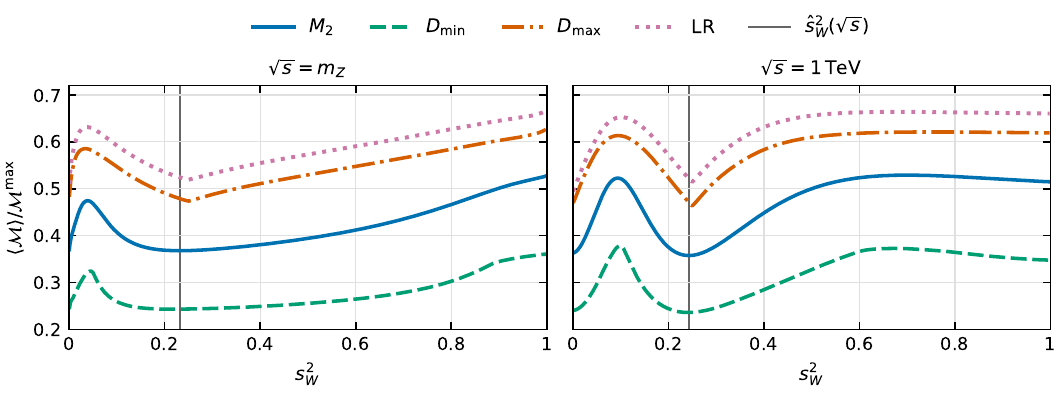}
  \caption{Average magic $\langle\mathcal{M}\rangle$ of the outgoing spin
  state in M{\o}ller scattering as a function of $s_W^2$, at
  $\sqrt{s}=m_Z$ (left) and $1\,\mathrm{TeV}$ (right), normalized to the
  two-qubit maximum of each measure. The gray vertical lines mark the SM
  $\overline{\mathrm{MS}}$ value $\hat{s}_W^2(\sqrt{s})$ at the
  corresponding energy.
}
  \label{fig:sW2}
\end{figure*}

How far apart are the sets of maximizers? Table~\ref{tab:dist} lists
their Fubini--Study distances~\cite{BengtssonZyczkowski2017}
$d(\psi,\psi')=\arccos|\braket{\psi|\psi'}|$.
The minimal distance $\pi/12$
between the $\mathfrak{D}$ and $M_2$ maximizers has a simple origin: the
$M_2$ maximizers lie on the boundary of the neighborhoods of radius
$\pi/12$ in which the $\mathfrak{D}$ maximizers are isolated. 
The LR-maximizing planes approach the
$\mathfrak{D}$ maximizers most closely but, being extended, also contain
states far from both other sets. Two Haar-random two-qubit states are on
average $11\pi/32\approx1.08$ apart~\cite{BengtssonZyczkowski2017}, so
the maximizers of different measures are close to each other. This,
however, only reflects that each set consists of Clifford orbits: a
Haar-random state lies on average $0.2185$ from the Clifford orbit of
another, comparable to the distances in Table~\ref{tab:dist}. The three
sets are visualized in App.~\ref{app:LR}.

\section{Magic and the weak mixing angle}
\label{sec:ew}

We now turn to a physics application in which the choice of measure
matters. In Ref.~\cite{Liu:2025bgw} it was found that the
weak mixing angle $s_W^2\equiv\sin^2\theta_W$ that minimizes the average
$M_2$ produced in M{\o}ller scattering, $e^-e^-\to e^-e^-$, agrees with
the measured $\overline{\mathrm{MS}}$ value
$\hat{s}_W^2(m_Z)=0.23122(6)$~\cite{ParticleDataGroup:2026mpi} at the sub-percent level and
tracks its renormalization-group running up to the TeV scale. We repeat
this analysis with $\mathfrak{D}_{\min}$, $\mathfrak{D}_{\max}$, and LR.

At tree level, M{\o}ller scattering proceeds through photon and $Z$
exchange in the $t$ and $u$ channels. The $Z$ couples to the electron
with the vertex structure
$\frac{e}{2s_Wc_W}\gamma^\mu(g_V-g_A\gamma^5)$, where
$c_W\equiv\cos\theta_W$, $g_A=-\frac12$, and $g_V=-\frac12+2s_W^2$. The
two qubits are the spins of the two outgoing electrons, quantized along
fixed axes in the center-of-mass frame: $\hat{z}$ along the incoming beam
and $\hat{y}$ normal to the scattering
plane~\cite{Liu:2025bgw}. The incoming spins are prepared in
one of the 60 two-qubit stabilizer states $\ket{\psi_{\mathrm{in},i}}$,
$i=1,\ldots,60$~\cite{Liu:2025qfl}. For each of them, the tree-level
amplitude produces an outgoing spin state
$\ket{\psi_{\mathrm{out},i}(\Theta;s_W^2,\sqrt{s})}$, normalized to unit
norm, where $\Theta$ is the polar scattering angle and $\sqrt{s}$ the
center-of-mass energy. Its magic, as quantified by a measure
$\mathcal{M}$, is averaged uniformly over the incoming stabilizer states
and over the solid angle $\Omega$ of the outgoing momentum,
\begin{equation}
  \langle\mathcal{M}\rangle(s_W^2;\sqrt{s})=
  \frac{1}{4\pi}\int d\Omega\,\frac{1}{60}\sum_{i=1}^{60}
  \mathcal{M}\big(\psi_{\mathrm{out},i}\big).
\end{equation}
Following Ref.~\cite{Liu:2025bgw}, we treat $s_W^2$ as a free
parameter, with $m_e$, $m_Z$, and $\Gamma_Z$ fixed to their PDG 2026
values~\cite{ParticleDataGroup:2026mpi}, and look for the value of
$s_W^2$ that minimizes $\langle\mathcal{M}\rangle$ at each $\sqrt{s}$.
Each measure is normalized by its two-qubit maximum from
Sec.~\ref{sec:maximal}.

Figure~\ref{fig:sW2} shows $\langle\mathcal{M}\rangle/\mathcal{M}^{\max}$
as a function of $s_W^2$ at $\sqrt{s}=m_Z$ and $1\,\mathrm{TeV}$. All
four curves have a peak at small $s_W^2$, followed by a minimum near
$s_W^2\approx0.2$--$0.25$, but the minima fall into two distinct
classes. For $M_2$ and $\mathfrak{D}_{\min}$, the minimum is smooth and
moves with the energy. At $\sqrt{s}=m_Z$, $M_2$ is minimized at
$s_W^2=0.2317$,
while the very
shallow $\mathfrak{D}_{\min}$ minimum lies $7\%$ below the measured
value. For $\mathfrak{D}_{\max}$ and LR, by contrast, the minimum is a
cusp located at exactly $s_W^2=1/4$.

\begin{figure}[t]
  \centering
  \includegraphics[width=\columnwidth]{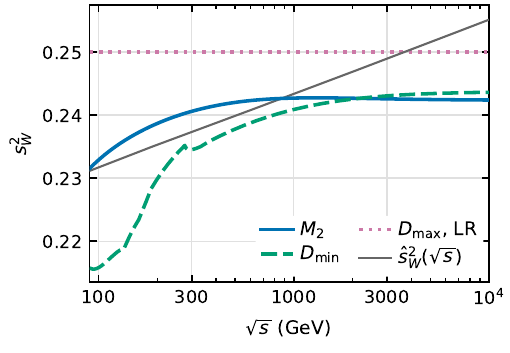}
  \caption{Value of $s_W^2$ minimizing the average magic in M{\o}ller
  scattering as a function of $\sqrt{s}$, for the four measures. The
  minima of $\mathfrak{D}_{\max}$ and LR both lie at $s_W^2=1/4$ at all
  energies (dotted line). For each measure and energy, the minimum is
  taken over $0.15\le s_W^2\le0.35$. The solid gray line is the SM running
  $\hat{s}_W^2(\sqrt{s})$ in the $\overline{\mathrm{MS}}$
  scheme~\cite{ParticleDataGroup:2026mpi,Erler:2004in,Erler:2017knj}.}
  \label{fig:minS}
\end{figure}

The two classes persist over the whole energy range, as
Fig.~\ref{fig:minS} shows, where the SM running $\hat{s}_W^2(\sqrt{s})$
is obtained from the $\overline{\mathrm{MS}}$ renormalization-group
equations of Refs.~\cite{Erler:2004in,Erler:2017knj}. The $M_2$ curve
tracks it up to about $1\,\mathrm{TeV}$, while the $\mathfrak{D}_{\min}$
curve rises more steeply. Its small kinks, most visibly near
$280\,\mathrm{GeV}$, reflect changes of the closest stabilizer state;
since the minimum of $\langle\mathfrak{D}_{\min}\rangle$ is very shallow,
they are sensitive to corrections beyond tree level and should not be
regarded as robust. Thus the same ``minimal magic'' hypothesis leads to
three different conclusions, depending on the measure: (i) with $M_2$,
it predicts the weak mixing angle at $m_Z$ and its running up to the TeV
scale; (ii) with $\mathfrak{D}_{\min}$, it predicts a similar value at
high energies but fails at $m_Z$ by $7\%$; and (iii) with
$\mathfrak{D}_{\max}$ or LR, it predicts the energy-independent value
$s_W^2=1/4$. This value also arises in the analysis of
$e^+e^-\to\mu^+\mu^-$ in Ref.~\cite{Cervera-Lierta:2017tdt} from
requiring maximal entanglement.

This pattern reflects the non-analytic structure of the measures.
$\mathfrak{D}_{\max}$ and LR are, by duality, maxima over witnesses, so
their kinks, where the optimal witness changes, always point upward and
can produce minima; $\mathfrak{D}_{\min}$ can only have downward kinks,
and $M_2$ has none. Since
$\langle\mathcal{M}\rangle$ is a linear average, an upward kink in the
contribution of a single incoming state already produces a cusp of the
average. The singlet $\ket{\Psi^-}=(\ket{01}-\ket{10})/\sqrt2$ is an
example. At $s_W^2=1/4$, the vector coupling $g_V=2(s_W^2-\frac14)$
vanishes, and the singlet is scattered into itself, a stabilizer state,
at every scattering angle. For a small $g_V\ne0$, its outgoing state acquires an
admixture linear in $g_V$. As for the single qubit near $\ket{0}$, this
increases $\mathfrak{D}_{\max}$ and LR by an amount $\propto|g_V|$, but
$M_2$ and $\mathfrak{D}_{\min}$ only by an amount $\propto g_V^2$; 
see App.~\ref{app:moller} for details. Other inputs, whose
outgoing states have finite magic at $g_V=0$, contribute upward kinks of
$\mathfrak{D}_{\max}$ and LR there as well.
The resulting cusp pins the minimum of
$\langle\mathfrak{D}_{\max}\rangle$ and $\langle\mathrm{LR}\rangle$ at
$s_W^2=1/4$, whereas for $M_2$ and $\mathfrak{D}_{\min}$ the minimum is
set by the smooth balance between photon and $Z$ exchange, which depends
on the energy. Like the kinks of the $\mathfrak{D}_{\min}$ curve, however,
this pinning rests on exact tree-level properties of the outgoing states
at $g_V=0$ and may not be robust against higher-order corrections.

\section{Conclusion and discussion}
\label{sec:conclusion}

We have compared four measures of magic, the SRE $M_2$, the min- and
max-relative entropies of magic $\mathfrak{D}_{\min}$ and
$\mathfrak{D}_{\max}$, and the logarithmic free robustness LR, for two
qubits. Although all four are maximized by the same $T$-type states for a
single qubit, they single out different maximal-magic states for two
qubits. $\mathfrak{D}_{\min}$ and $\mathfrak{D}_{\max}$ share the same
maximizers, a consequence of their duality, which are distinct from
those of $M_2$ and from the planes of LR maximizers. We have provided
exact proofs of these statements, and the approaches may be generalized
to test the conjectured maximal-magic states of more
qubits~\cite{Liu:2025frx} and of qudits. The maximizers of
$\mathfrak{D}_{\min}$ and $M_2$ are nearly maximal in all measures,
whereas the LR maximizers can be far from maximal in $M_2$ and
$\mathfrak{D}_{\min}$.

The measures also differ in their analytic structure:
$\mathfrak{D}_{\max}$ and LR grow linearly away from stabilizer states,
where $M_2$ and $\mathfrak{D}_{\min}$ grow only quadratically, and their
kinks always point upward. In M{\o}ller scattering, this makes them
predict $s_W^2=1/4$ at all energies, where the vector coupling of the $Z$
boson to charged leptons vanishes, whereas $M_2$ and
$\mathfrak{D}_{\min}$ predict an energy-dependent $s_W^2$, which for $M_2$
agrees with the measured value and its running up to the TeV scale.

These results show that statements about magic in physical processes,
such as whether a given interaction minimizes or maximizes magic, may
depend on the measure. This does not diminish the interest of the
results obtained with the SRE; rather, it raises the question of which
measure, if any, is singled out by the physics at hand, and why.
$\mathfrak{D}_{\min}$, $\mathfrak{D}_{\max}$, and LR are operationally
linked to stabilizer fidelity, stabilizer decompositions, and
quasiprobability simulation, respectively, while the SREs are linked to
the flatness of the Pauli spectrum, to stabilizer property testing, and
to how closely Clifford orbits mimic Haar-random
states~\cite{Bittel:2025yhq}. A principle of ``minimal quantum
computational resources'' in nature would have to specify which of
these resources is minimized. It is therefore important to study
different magic measures in more detail and to consider measures beyond
the SRE when applying the concept of magic to high-energy physics.

\begin{acknowledgments}
We thank Sokratis Trifinopoulos for helpful comments on the manuscript. Z. Y. thanks Kun Cheng, Ying-Ying Li, Qiaofeng Liu, Ian Low and Yi-Lin Wang for useful discussions. Z.-H. S. and H.-L. Z. thank USTC for its support of the summer internship program at IHEP from July 13 to 24, 2026, during which part of this work was carried out. Z. Y. is supported by IHEP under Grant No. E5515AU1.
\end{acknowledgments}

\appendix
\section{Clifford gates and the reduced search domain}
\label{app:domain}

In the computational basis $\{\ket{0},\ket{1}\}$, the Pauli matrices are
\begin{equation}
  X=\begin{pmatrix}0&1\\1&0\end{pmatrix},\quad
  Y=\begin{pmatrix}0&-i\\i&0\end{pmatrix},\quad
  Z=\begin{pmatrix}1&0\\0&-1\end{pmatrix},
\end{equation}
and the single-qubit Clifford generators of Sec.~\ref{sec:measures} are
the Hadamard and phase gates
\begin{equation}
  H=\frac{1}{\sqrt2}\begin{pmatrix}1&1\\1&-1\end{pmatrix},\qquad
  S=\begin{pmatrix}1&0\\0&i\end{pmatrix}.
\end{equation}
For two qubits we use the ordered basis
$\{\ket{00},\ket{01},\ket{10},\ket{11}\}$, in which the CNOT gate with
the first qubit as control and the second as target is
\begin{equation}
  \mathrm{CNOT}_{1\to2}=\begin{pmatrix}
  1&0&0&0\\0&1&0&0\\0&0&0&1\\0&0&1&0\end{pmatrix}.
\end{equation}
Up to a global phase, a two-qubit state
$\ket{\psi}=c_1\ket{00}+c_2\ket{01}+c_3\ket{10}+c_4\ket{11}$ is
parameterized as~\cite{Liu:2025frx}
\begin{equation}
\begin{aligned}
  c_1&=\sin\theta_1\sin\theta_2\,e^{i\phi_1}, &
  c_2&=\sin\theta_1\cos\theta_2\,e^{i\phi_2},\\
  c_3&=\cos\theta_1\sin\theta_3\,e^{i\phi_3}, &
  c_4&=\cos\theta_1\cos\theta_3,
\end{aligned}
\label{eq:param}
\end{equation}
with $\theta_k\in[0,\pi/2]$ and $\phi_k\in[0,2\pi)$. The search of
Sec.~\ref{sec:maximal} uses the reduced domain
\begin{equation}
\begin{gathered}
  \theta_1=u_1\arctan\frac{\sin\theta_3}{\cos\theta_2},\quad
  \theta_2=\frac{\pi}{4}u_2,\quad \theta_3=\frac{\pi}{4}u_3,\\
  \phi_1=\frac{\pi}{2}(u_4+u_5),\quad \phi_2=\frac{\pi}{2}u_4,\quad
  \phi_3=\frac{\pi}{2}(u_5+2u_6),
\end{gathered}
\label{eq:domain}
\end{equation}
with $u_k\in[0,1]$, and every two-qubit state is Clifford-equivalent to
at least one state in it. With these gates we now derive the
domain~\eqref{eq:domain}, using that
all four measures are invariant under Clifford unitaries. We first
generate all permutations of the amplitudes and then restrict the relative
phases.

$\mathrm{CNOT}_{1\to2}$ exchanges
$\ket{10}\leftrightarrow\ket{11}$, i.e., $c_3\leftrightarrow c_4$.
$\mathrm{CNOT}_{2\to1}=(H\otimes H)\,\mathrm{CNOT}_{1\to2}\,(H\otimes H)$
exchanges $c_2\leftrightarrow c_4$, and
$\mathrm{SWAP}=\mathrm{CNOT}_{1\to2}\mathrm{CNOT}_{2\to1}\mathrm{CNOT}_{1\to2}$
exchanges $c_2\leftrightarrow c_3$. These transpositions generate all
permutations of $\{c_2,c_3,c_4\}$. Finally, $X\otimes I=(HS^2H)\otimes I$
acts as $(c_1\,c_3)(c_2\,c_4)$, so together with the above it generates
the full permutation group of $\{c_1,c_2,c_3,c_4\}$. We can therefore
impose $|c_1|\le|c_2|\le|c_3|\le|c_4|$. In the
parameterization~\eqref{eq:param}, $|c_1|\le|c_2|$ and
$|c_3|\le|c_4|$ give $\theta_2,\theta_3\le\pi/4$, and $|c_2|\le|c_3|$ gives
$\tan\theta_1\le\sin\theta_3/\cos\theta_2$.

Next, the diagonal Clifford unitaries $S\otimes I$,
$I\otimes S$, and $CZ=(I\otimes H)\,\mathrm{CNOT}_{1\to2}\,(I\otimes H)$
leave $|c_k|$ unchanged. Up to global phases, the combinations
\begin{equation}
\begin{aligned}
  {}[-i(S\otimes I)]^3&=\mathrm{diag}(i,i,1,1),\\
  {}[-i(I\otimes S)]^3&=\mathrm{diag}(i,1,i,1),\\
  (-CZ)\,[-i(S\otimes I)]^2&=\mathrm{diag}(1,1,-1,1)
\end{aligned}
\end{equation}
shift the relative phases $(\phi_1,\phi_2,\phi_3)$ by the lattice vectors
$\frac{\pi}{2}(1,1,0)$, $\frac{\pi}{2}(1,0,1)$, and $\pi(0,0,1)$. A unit
cell of this lattice is
$(\phi_1,\phi_2,\phi_3)=(\beta_1+\beta_2,\beta_1,\beta_2+2\beta_3)$ with
$\beta_k\in[0,\pi/2]$, which has volume $\pi^3/4=(2\pi)^3/32$. Combined
with the factor $1/4!$ from the ordering of the moduli, the
domain~\eqref{eq:domain} covers $1/768$ of the original parameter
space. We note that this reduction does not make the domain small in
state space: Fubini--Study distances between states in it can still reach
$\pi/2$.

\section{Single-qubit measures in closed form}
\label{app:1qubit}

For the single-qubit state
$\ket{\psi}=\cos\frac{\theta}{2}\ket{0}+e^{i\varphi}\sin\frac{\theta}{2}\ket{1}$
of Sec.~\ref{sec:measures}, with Bloch vector
$\bm{r}=(x,y,z)=(\sin\theta\cos\varphi,\,\sin\theta\sin\varphi,\,\cos\theta)$,
the stabilizer polytope is the octahedron $|x|+|y|+|z|\le1$. Let
$r_{(1)}\ge r_{(2)}\ge r_{(3)}$ denote $|x|$, $|y|$, and $|z|$ in
decreasing order, so that $\|\bm{r}\|_\infty=r_{(1)}$ and
$\|\bm{r}\|_1=r_{(1)}+r_{(2)}+r_{(3)}$. One finds
\begin{equation}
\begin{aligned}
  M_2 &= -\log_2\frac{1+x^4+y^4+z^4}{2}, \\
  \mathfrak{D}_{\min} &= -\log_2\frac{1+\|\bm{r}\|_\infty}{2}, \qquad
  \mathrm{LR} = \log_2\frac{1+\|\bm{r}\|_1}{2},
\end{aligned}
\label{eq:1qubit}
\end{equation}
and
\begin{equation}
  2^{\mathfrak{D}_{\max}}=
  \begin{cases}
    1+\frac{\sqrt3-1}{2}\big(\|\bm{r}\|_1-1\big), & r_{(3)}>\eta,\\[3pt]
    \big(\sqrt{1-r_{(1)}}+\sqrt{1-r_{(2)}}\,\big)^2, & r_{(3)}\le\eta,
  \end{cases}
  \label{eq:1qubitDmax}
\end{equation}
where $\eta=\tan\frac{\pi}{12}\,(r_{(1)}+r_{(2)}-1)$ and
$\tan\frac{\pi}{12}=2-\sqrt3$. The first case covers a region around
each $T$-type state, including the whole path of Fig.~\ref{fig:1qubit}.
The second covers bands around the great circles through the
stabilizer states, on which $r_{(3)}=0$. In the first region the optimal
stabilizer decomposition uses the three stabilizer states closest to
$\ket{\psi}$, and $2^{\mathfrak{D}_{\max}}-1=(\sqrt3-1)(2^{\mathrm{LR}}-1)$.
In the second it uses only the two closest ones.
The single-qubit $\mathfrak{D}_{\min}$ in Eq.~\eqref{eq:1qubit} agrees
with Ref.~\cite{Rubboli:2023qzz}, and
Eq.~\eqref{eq:1qubitDmax} is the pure-state case of the closed form
of $2^{\mathfrak{D}_{\max}}$, the generalized robustness of magic,
for all single-qubit
states derived there. Equations~\eqref{eq:1qubit} and
\eqref{eq:1qubitDmax} show that all four measures are maximized by the
$T$-type states, $|x|=|y|=|z|=1/\sqrt{3}$.

Besides the conical non-analyticity at the stabilizer states discussed in
Sec.~\ref{sec:measures}, $\mathfrak{D}_{\max}$ is non-analytic along the
boundaries between the two cases of Eq.~\eqref{eq:1qubitDmax}, where the
optimal decomposition switches from three to two stabilizer states. There
it is continuously differentiable, but its second derivatives jump.

\section{Global maximality of the $M_2$ maximizers}
\label{app:M2proof}

Here we prove that $M_2\le\log_2(16/7)$ for all two-qubit pure states,
with equality only on the 480 states of Ref.~\cite{Liu:2025frx}.
The vector $m$, the matrix $M$, and
the 480 maximizers are given as data files in the Supplemental
Material~\cite{SM}.

\emph{From a maximum to a positive polynomial.}---Let
$T_P=\bra{\psi}P\ket{\psi}$ for the 16 Pauli strings $P$, with
$T_{II}=\braket{\psi|\psi}$. For normalized states,
Eq.~\eqref{eq:M2} gives $2^{-M_2}=\frac14\sum_PT_P^4$, so the claim is
$\frac14\sum_PT_P^4\ge\frac7{16}$ on the unit sphere. Multiplying by
powers of $T_{II}$ removes the normalization condition, and the claim
becomes
\begin{equation}
  G(\psi)\equiv\frac14\sum_PT_P^4\,T_{II}^4-\frac7{16}\,T_{II}^8\ \ge\ 0
  \qquad\text{for all }\psi\in\mathbb{C}^4 .
  \label{eq:Fpos}
\end{equation}
$G$ is a real homogeneous polynomial of degree 8 in $\psi$ and 8 in
$\psi^*$. The statement that $M_2^{\max}=\log_2(16/7)$ is a global maximum
is thus equivalent to the statement that $G$ is positive semidefinite,
i.e., non-negative everywhere.
A standard sufficient condition is a sum-of-squares (SOS)
representation~\cite{Parrilo2003,JeanB:2006win}
\begin{equation}
  G=m^{T}M\,m,\qquad M\succeq0,
  \label{eq:SOS}
\end{equation}
where $m$ is a vector of real polynomials and $M$ is a constant real
symmetric matrix. Diagonalizing
$M=\sum_k\mu_kv_kv_k^{T}$ with $\mu_k\ge0$ gives
$G=\sum_k\mu_k(v_k\cdot m)^2\ge0$ manifestly. Once $m$ is chosen,
Eq.~\eqref{eq:SOS} is linear in $M$, so finding $M$ is a semidefinite
program.

\emph{Pauli variables.}---The natural variables are the real
expectation values $T_P$, on which Clifford unitaries act by signed
permutations. We take $m$ to be monomials of degree up to 4 in the 15
nontrivial $T_P$, multiplied by powers of $T_{II}$ to make each of them
of degree 4. The key point is that Eq.~\eqref{eq:SOS} has to hold only
on pure states, i.e., as an identity of polynomials in $\psi$ and
$\psi^*$, not for independent $T_P$. Regarded as a function of
independent variables, $G$ is negative, e.g., at $T_P=0$ for
$P\ne II$. It is the quadratic relations among the $T_P$ of a pure state,
such as $\sum_PT_P^2=4T_{II}^2$, that make a certificate possible. We
find that no certificate of this type exists at degree 6, while at
degree 8 a small subset of monomials suffices. These are the monomials
whose Pauli strings form a single string, a commuting triple
$\{P,Q,PQ\}$, or an anticommuting pair, 751 in total.

\emph{Constraints on $M$.}---Two features fix the structure of $M$.
First, $G$ vanishes on the 480 maximizers $\psi_j$. Since $M\succeq0$,
this forces $M\,m(\psi_j)=0$. The 480 vectors $m(\psi_j)$ span a
225-dimensional subspace, so $M$ necessarily has 225 zero eigenvalues.
Second, $G$ is
Clifford invariant, so $M$ can be taken Clifford invariant as well. This
leaves 140 free parameters. A semidefinite program finds an $M$ that is
strictly positive on the complement of the forced kernel. Its smallest
eigenvalue there is about $0.016$. Because of this margin, we can round
the parameters to rationals and solve the linear constraints exactly
without losing positivity. The resulting $M$ has rational entries whose
denominators divide $3\cdot2^{47}$.

\emph{Exact verification.}---Two facts are then checked in exact
arithmetic. (i)~Eq.~\eqref{eq:SOS} holds identically: both sides are
expanded as polynomials in $\psi$ and $\psi^*$, and all coefficients
agree. (ii)~$M\succeq0$: $M$ annihilates the 225-dimensional subspace,
and on its 526-dimensional complement it is related by an exact integer
congruence to a strictly diagonally dominant matrix, which is positive
definite. Together they prove Eq.~\eqref{eq:Fpos}, i.e.,
$M_2\le\log_2(16/7)$.

\emph{Uniqueness of the maximizers.}---Step (ii) shows more: the kernel
of $M$ is \emph{exactly} the span of the 480 vectors $m(\psi_j)$. If
$G(\psi)=0$ for some state, then $M\,m(\psi)=0$, so $m(\psi)$ lies in
this span. Consequently, every polynomial $\gamma\cdot m$ that vanishes at
the 480 known maximizers also vanishes at any maximizer, known or not.
The choice $\gamma\cdot m=T_{II}T_P(4T_P^2-T_{II}^2)$ vanishes at the 480
states, since all their $T_P$ lie in $\{0,\pm\frac12\}$. It therefore
forces $T_P\in\{0,\pm\frac12\}$ for every maximizer and every $P$. This
leaves a finite problem: among the $3^{15}$ candidate Pauli vectors, an
exhaustive exact test of the purity condition
$\rho^2=\rho$, with $\rho=\frac14\sum_PT_PP$, finds exactly 480
solutions, which are the known maximizers. The argument does not assume
that the numerical search of Ref.~\cite{Liu:2025frx} found all
maximizers.

The same strategy applies to other numbers of qubits and to qudits: an
SOS certificate in the Pauli expectation values, with the kernel fixed by
the conjectured maximizers.

\section{Global maximality of the $\mathfrak{D}$ maximizers}
\label{app:Dproof}

Here we prove that $F_{\mathrm{STAB}}(\psi)\ge F_*\equiv(2+\sqrt3)/6$ for all
two-qubit pure states, with equality exactly on the 640 Clifford images of
$\ket{\psi_{\mathfrak{D}}}$ of Eq.~\eqref{eq:psiD}. By the duality
discussed in Sec.~\ref{sec:maximal}, this settles the maxima and the
maximizers of both $\mathfrak{D}_{\min}$ and $\mathfrak{D}_{\max}$. Unlike
$2^{-M_2}$ in App.~\ref{app:M2proof}, $F_{\mathrm{STAB}}$ is not a
polynomial. It is the maximum of the 60 quadratic functions
$f_s(\psi)=|\braket{s|\psi}|^2$, and its minima are cusps, as for a
single qubit (Fig.~\ref{fig:1qubit}). We therefore combine an exact local
analysis at the maxima with a global branch-and-bound search.

\emph{Local analysis.}---Write a neighboring state as
$\ket{\psi}=(\ket{\psi_{\mathfrak{D}}}+\ket{\delta})/\sqrt{1+\|\delta\|^2}$
with $\braket{\psi_{\mathfrak{D}}|\delta}=0$, so that
$\|\delta\|=\tan d$, where $d$ is the Fubini--Study distance. At
$\psi_{\mathfrak{D}}$ the maximal overlap $F_*$ is attained by nine
\emph{active} stabilizer states, six with four nonzero amplitudes and three
with two, e.g., $\frac12(\ket{00}+\ket{01}+\ket{10}+\ket{11})$ and
$\frac{1}{\sqrt2}(\ket{01}+\ket{11})$. All other
overlaps are at most $1/2$. For an active $s$ one has exactly
\begin{equation}
\begin{aligned}
  f_s(\psi)&=\frac{F_*+h_s(\delta)+|\braket{s|\delta}|^2}{1+\|\delta\|^2},\\
  h_s(\delta)&=2\,\mathrm{Re}\big[\braket{\psi_{\mathfrak{D}}|s}\braket{s|\delta}\big],
\end{aligned}
\end{equation}
where $h_s$ is the linear term of the first-order analysis of
Ref.~\cite{Erew:2025egp}. Regarded as vectors in the six-dimensional
real tangent space, the nine $h_s$ span this space and sum to zero.
The origin is therefore an interior point of the polytope
$\mathcal{K}=\mathrm{conv}\{h_s\}$, and
$\max_sh_s(\delta)\ge\kappa\|\delta\|$, where $\kappa$ is the inradius of
$\mathcal{K}$. An exact enumeration of its 27 facets gives $\kappa=1/6$.
Hence
\begin{equation}
  F_{\mathrm{STAB}}(\psi)-F_*\ \ge\
  \frac{\|\delta\|\,(\frac16-F_*\|\delta\|)}{1+\|\delta\|^2}\ >\ 0
\end{equation}
for $0<\|\delta\|<1/(6F_*)$, i.e., for $0<d<\pi/12$, since
$36F_*^2/(36F_*^2+1)=(2+\sqrt3)/4=\cos^2(\pi/12)$. Each maximizer is thus a
strict, cusp-like local maximum of $\mathfrak{D}_{\min}$, and the only
maximizer within distance $\pi/12$. By Clifford invariance, the same holds
for all 640 states.

\emph{Global search.}---It remains to show that $F_{\mathrm{STAB}}>F_*$
outside these balls. Using the symmetry reduction of App.~\ref{app:domain},
written in the affine chart $\tilde{\bm c}=(c_1,c_2,c_3)/c_4$, we cover the
reduced domain by boxes. Since each $|\braket{s|\psi}|$ is 1-Lipschitz,
$\sqrt{F_{\mathrm{STAB}}(\psi)}\ge\sqrt{F_{\mathrm{STAB}}(\psi_b)}-r_b$
for every state $\psi$ in a box with center $\psi_b$, where $r_b$ bounds
the distance between the normalized states, minimized over the global
phase. If $\ell_b$ is the half-diagonal of the box and
$1+\|\tilde{\bm c}\|^2\ge\omega_b^2$ on it, the unnormalized vectors
$(\tilde{\bm c},1)$ differ from that of the center by at most $\ell_b$,
so the Fubini--Study distance $d$ to the center obeys
$\sin d\le\ell_b/\omega_b$. The distance $\sqrt{2(1-\cos d)}$ is then at
most $r_b\equiv\sqrt{2\ell_b^2/(2\omega_b^2-\ell_b^2)}$. Every box is either
certified, $\sqrt{F_{\mathrm{STAB}}(\psi_b)}-r_b>\sqrt{F_*}$, or shown to
lie inside one of the balls of radius $\pi/12$. Box centers are dyadic
rationals, so $F_{\mathrm{STAB}}(\psi_b)$ is an exact rational number and
both tests reduce to integer inequalities. Floating-point arithmetic is
used only to choose which test to apply. About $4.4\times10^6$ boxes
suffice.

The most delicate region surrounds the 480 maximizers of $M_2$. They are
local minima of $F_{\mathrm{STAB}}$ with value $5/8$, only slightly above
$F_*$, and lie at Fubini--Study distance exactly $\pi/12$ from the nearest
$\mathfrak{D}$ maximizer, on the boundary of the balls above. This is the
origin of the minimal distance $\pi/12$ in Table~\ref{tab:dist}.

\section{Global maximality of the LR maximizers}
\label{app:LRproof}

Here we prove that the robustness of magic satisfies
$\mathcal{R}(\rho)\le\sqrt5$, i.e., $\mathrm{LR}(\rho)\le\log_2\frac{1+\sqrt5}{2}$,
for all two-qubit states, and we determine the states that attain the
bound. Unlike for $M_2$ and $\mathfrak{D}_{\min}$, linear-programming duality
turns the problem into a finite one.

\emph{Reduction to a finite problem.}---$\mathcal{R}(\rho)$ is the minimal
$\ell_1$ norm of a real decomposition of $\rho$ into the 60 stabilizer
projectors. Its dual linear program reads
\begin{equation}
\begin{aligned}
  \mathcal{R}(\rho)=\max\big\{&\mathrm{Tr}(A\rho):\ A=A^\dagger,\\
  &|\braket{s|A|s}|\le1\ \text{for all }s\in\mathrm{STAB}_2\big\}.
\end{aligned}
\label{eq:Rdual}
\end{equation}
Write $A=\sum_Pa_PP$ in the Pauli basis. The four stabilizer states with a
given stabilizer group $\{I,P,Q,R\}$ diagonalize $a_II+a_PP+a_QQ+a_RR$, so the
constraints say that this operator has norm at most 1, for each of the 15
such commuting triples. The feasible set $\mathcal{A}$ is a bounded polytope
in $\mathbb{R}^{16}$ with 120 facets.
Its vertices are rational, because the
constraint coefficients are $0$ or $\pm1$. The maximum of a linear function
over $\mathcal{A}$ is attained at a vertex, so
\begin{equation}
  \max_\rho\mathcal{R}(\rho)=\max_{v}\lambda_{\max}(A_v),
\end{equation}
where $v$ runs over the vertices of $\mathcal{A}$, $A_v$ is the operator
at vertex $v$, and $\lambda_{\max}(A_v)$ is its largest eigenvalue, i.e.,
the maximum of $\mathrm{Tr}(A_v\rho)$ over all states $\rho$.

\emph{The extremal operators.}---There are exactly six sets of five
mutually anticommuting non-identity Pauli strings, and any commuting triple
contains exactly one element of each set. For such a set $\{P_k\}$, the
operator $A^*=\sum_k\epsilon_kP_k$, $\epsilon_k=\pm1$, therefore satisfies
$\braket{s|A^*|s}=\pm1$ for all 60 stabilizer states, so $A^*\in\mathcal{A}$.
Since the $P_k$ anticommute, $(A^*)^2=5I$, and the eigenvalue $\sqrt5$ of
$A^*$ is doubly degenerate. The $6\times2^5=192$ operators $A^*$ form a
single Clifford orbit, and their top eigenspaces are the 192 planes of
Sec.~\ref{sec:maximal}. For example, the plane spanned by
Eq.~\eqref{eq:LRplane} belongs to $XI-YX-YY+YZ-ZI$.

\emph{Vertex enumeration.}---We enumerate all vertices of $\mathcal{A}$
exactly, up to the Clifford group and $A\to-A$. We use adjacency
decomposition: starting from $A^*$, the neighbors of each new vertex orbit
are found from the extreme rays of its tangent cone, computed by an exact
double-description method in integer arithmetic. The search is complete
because the graph of a polytope is connected. It yields 626\,626 vertices in
44 orbits.

\emph{Exact bound.}---For each vertex orbit, the characteristic polynomial
$p$ of $A_v$ is real-rooted. Hence $\lambda_{\max}(\pm A_v)\le\sqrt5$, i.e.,
the spectrum of $A_v$ lies in $[-\sqrt5,\sqrt5]$, if and only if all
coefficients of $p(w+\sqrt5)$ and of $p(-w-\sqrt5)$ are non-negative, which
we check exactly in $\mathbb{Q}(\sqrt5)$. Both signs are needed because the
orbits are taken under $A\to-A$, which is a symmetry of the feasible set but
not of the spectrum. All 44 orbits pass. The value $\sqrt5$ is attained only
by the orbit of $A^*$. The next largest eigenvalue,
$(1+2\sqrt3)/2\approx2.232$, is the robustness of a pair of $T$-type
states~\cite{Howard:2017maw}. If $\mathcal{R}(\rho)=\sqrt5$, an optimal
vertex $v$ satisfies $\mathrm{Tr}(A_v\rho)=\sqrt5\ge\lambda_{\max}(A_v)$. Then
$v$ is one of the operators $A^*$, and $\rho$ is supported on its top
eigenspace. Conversely, every such state attains
$\mathcal{R}=\mathrm{Tr}(A^*\rho)=\sqrt5$. The maximizers of LR are
therefore exactly the states supported on one of the 192 planes, including
mixed ones.

\section{The LR-maximizing planes}
\label{app:LR}

The 192 LR-maximizing planes are two-dimensional subspaces. The plane spanned by Eq.~\eqref{eq:LRplane}
is the solution space of
\begin{equation}
\begin{aligned}
  -(1+\sqrt5)c_1+(1-i)c_3+(1+i)c_4&=0,\\
  -(1+\sqrt5)c_2-(1-i)c_3+(1+i)c_4&=0,
\end{aligned}
\end{equation}
and intersects the interior of the domain~\eqref{eq:domain}, e.g., at
$(u_1,\ldots,u_6)\approx(0.847,0.812,0.844,0.154,0.837,0.678)$. The
pure states in each plane form a Bloch sphere. Consequently, for states
sampled uniformly on a plane, the mean and the median of the
Fubini--Study distance are both $\pi/4$, compared with $11\pi/32$ for
Haar-random states of the full two-qubit space.

Comparing the 192 planes pairwise, we find 2880 pairs that intersect,
each at a single state. These 2880 intersection states are all distinct,
so no state lies on three or more planes. Their concurrences take four
values, listed in Table~\ref{tab:LRint}. Each plane contains exactly two
product states, giving 384 distinct product states in total. On each
plane, $M_2=\log_2(25/12)$ is constant, whereas the other measures vary
within
\begin{equation}
\begin{aligned}
  \log_2\frac{10}{5+\sqrt5}&\le\mathfrak{D}_{\min}\le\mathfrak{D}_*,\\
  \mathfrak{D}_*&\le\mathfrak{D}_{\max}\le
    \log_2\Big(6-3\sqrt3+\frac{\sqrt{15}}{5}\Big),
\end{aligned}
\end{equation}
where
$\mathfrak{D}_*=\log_2\big[20/\big(5+\sqrt5+2\sqrt{5+2\sqrt5}\big)\big]$,
i.e., $0.4667\le\mathfrak{D}_{\min}\le0.5787\le\mathfrak{D}_{\max}\le0.6585$.
All four bounds are attained. The concurrence varies
from 0 to $\sqrt{3/5}$.

\begin{table}[h]
  \caption{Concurrence of the 2880 states at which two LR-maximizing
  planes intersect.}
  \label{tab:LRint}
  \begin{ruledtabular}
  \begin{tabular}{cc}
    Concurrence & Number of states \\
    \hline
    $\sqrt{(5-\sqrt5)/10}$ & 576 \\
    $\sqrt{2/5}$ & 1152 \\
    $\sqrt{3/5}$ & 576 \\
    $\sqrt{(5+\sqrt5)/10}$ & 576 \\
  \end{tabular}
  \end{ruledtabular}
\end{table}

\begin{figure*}[t]
  \centering
  \includegraphics[width=0.9\textwidth]{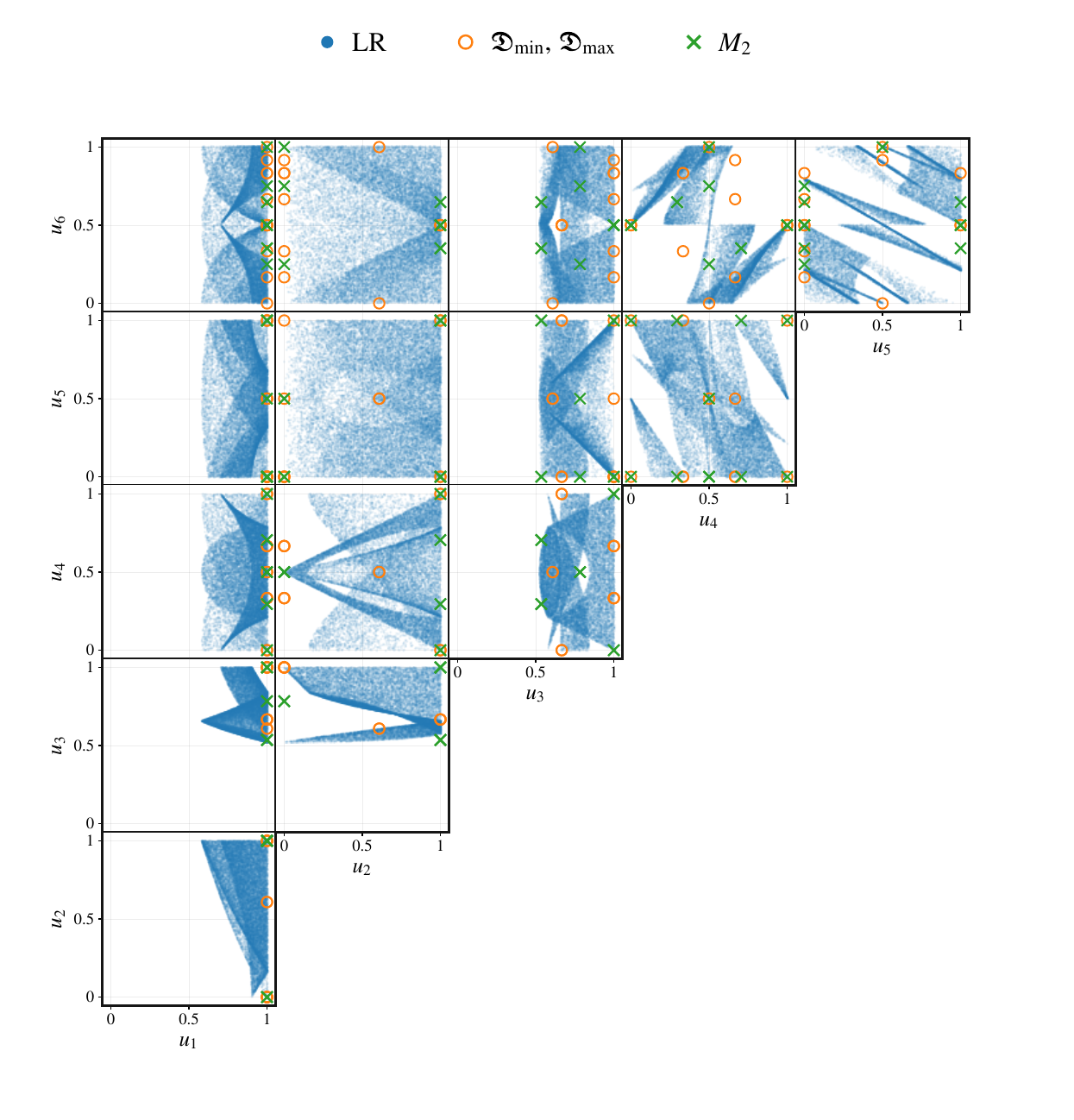}
  \caption{Two-parameter projections of the maximizers in the reduced
  domain~\eqref{eq:domain}: states sampled from the LR-maximizing
  planes (blue dots), the maximizers of $\mathfrak{D}_{\min}$ and
  $\mathfrak{D}_{\max}$ (orange open circles), and the maximizers of $M_2$
  (green crosses).}
  \label{fig:projection}
\end{figure*}

Figure~\ref{fig:projection} shows the locations of the three sets of
maximizers in the domain~\eqref{eq:domain}, projected onto each pair of
the parameters $u_k$. The LR maximizers fill extended regions, while the
$\mathfrak{D}$ and $M_2$ maximizers are isolated points, many of which lie
on the boundary of the domain. We stress that distances in these
projections are not faithful measures of distances in state space; the
quantitative comparison is given in Table~\ref{tab:dist}.

\section{Magic near $g_V=0$ in M{\o}ller scattering}
\label{app:moller}

Here we give the details of the discussion of the cusp at $s_W^2=1/4$ in
Sec.~\ref{sec:ew}. We first explain why the kinks of
$\mathfrak{D}_{\max}$ and LR point upward and those of
$\mathfrak{D}_{\min}$ downward, then describe how the measures grow near
a stabilizer state, and finally apply this to the singlet input near
$g_V=0$. All results are at tree level. The cusp relies on outgoing
states that are exactly stabilizer states at $g_V=0$, as for the singlet,
and on exact coincidences of optimal witnesses for the other inputs.
Higher-order corrections need not preserve these, and may therefore
shift the cusp away from $s_W^2=1/4$ or smooth it out.

\emph{Direction of kinks.}---Consider normalized states $\ket{\psi(g)}$
that depend smoothly on a real parameter $g$, such as an outgoing state
$\ket{\psi_{\mathrm{out},i}}$ at fixed $\Theta$ as a function of $s_W^2$.
By duality, $\sqrt{\xi}$ is a maximum over witnesses~\cite{Regula:2017mvm},
\begin{equation}
  \sqrt{\xi(\psi)}=\max\big\{|\braket{y|\psi}|:\ |\braket{s|y}|\le1\
  \ \forall s\in\mathrm{STAB}_n\big\},
  \label{eq:xidual}
\end{equation}
and so is the robustness $\mathcal{R}$, Eq.~\eqref{eq:Rdual}. The
stabilizer fidelity $F_{\mathrm{STAB}}$ is a maximum over the finitely
many overlaps $|\braket{s|\psi}|^2$. In each case the maximum is over a
compact set, of functions that are linear in the witness and as smooth in
$g$ as $\ket{\psi(g)}$; for $\sqrt\xi$ we use $\mathrm{Re}\braket{y|\psi}$,
which has the same maximum because the witness set is invariant under
phases. By Danskin's theorem~\cite{Danskin1966}, such a maximum $f(g)$ has one-sided
derivatives, which are the largest and the smallest derivative of the
optimal witnesses at $g$, so that $f'(g^-)\le f'(g^+)$. A kink, which can
occur only where the optimal witness changes, therefore always points
upward, like that of $|g|$. Since the logarithm is increasing, the same
holds for $\mathfrak{D}_{\max}=2\log_2\sqrt{\xi}$ and
$\mathrm{LR}=\log_2[(1+\mathcal{R})/2]$, whereas the kinks of
$\mathfrak{D}_{\min}=-\log_2F_{\mathrm{STAB}}$ point downward. $M_2$ is a
smooth function of the state, Eq.~\eqref{eq:M2}, and has no kinks. Upward
kinks of several contributions at the same $g$ add up to an upward kink
of their sum, and hence of the average $\langle\mathcal{M}\rangle$. This
extends to the integral over $\Theta$, because the derivatives of the
outgoing states with respect to $g$ are bounded uniformly in $\Theta$.
Their only potential obstructions are removed by rescaling: the photon
poles at $\Theta=0,\pi$, and the zero at $\Theta=\pi/2$ of the amplitude
for the input $(\ket{00}+\ket{11})/\sqrt2$, which is present for every
$g$. We have checked numerically that the rescaled amplitudes do not
vanish.

\emph{Onset near a stabilizer state.}---Let $\ket{s_0}$ be a pure
stabilizer state of $n$ qubits, let $\ket{\chi}$ be a unit vector with
$\braket{s_0|\chi}=0$, and consider
\begin{equation}
  \ket{\psi(\zeta)}=\frac{\ket{s_0}+\zeta\ket{\chi}}{\sqrt{1+\zeta^2}},\qquad
  \zeta\in\mathbb{R}.
\end{equation}
Then
\begin{equation}
  \mathfrak{D}_{\min}=\frac{\zeta^2}{\ln2}+O(\zeta^4),\qquad
  M_2=\frac{4\zeta^2}{\ln2}+O(\zeta^4),
  \label{eq:quadonset}
\end{equation}
whereas $\mathfrak{D}_{\max}$ and LR are of order $|\zeta|$: for
sufficiently small $|\zeta|$, both are bounded above and below by
positive multiples of $|\zeta|$.
As functions of $\zeta$, $\mathfrak{D}_{\max}$ and LR thus have an upward
kink at the stabilizer state, while $M_2$ and $\mathfrak{D}_{\min}$ are
smooth there.

\emph{The singlet input near $g_V=0$.}---We write $g_V=2(s_W^2-\frac14)$
and use the spin basis of Sec.~\ref{sec:ew}, and we define the Bell states
\begin{equation}
\begin{aligned}
  \ket{\Phi^\pm}&=\frac{\ket{00}\pm\ket{11}}{\sqrt2}, &
  \ket{\Psi^\pm}&=\frac{\ket{01}\pm\ket{10}}{\sqrt2}.
\end{aligned}
\end{equation}
At $g_V=0$, exactly two of the 60 incoming stabilizer states produce
stabilizer outgoing states at every scattering angle at which the
amplitude is nonzero, namely $\ket{\Phi^+}$ and $\ket{\Psi^-}$. For
$\ket{\Phi^+}$ the outgoing state is a stabilizer state for every value
of $g_V$, so this input does not produce magic at all. The singlet
$\ket{\Psi^-}$, in contrast, produces magic for every $g_V\ne0$.

\begin{figure}[t]
  \centering
  \includegraphics[width=\columnwidth]{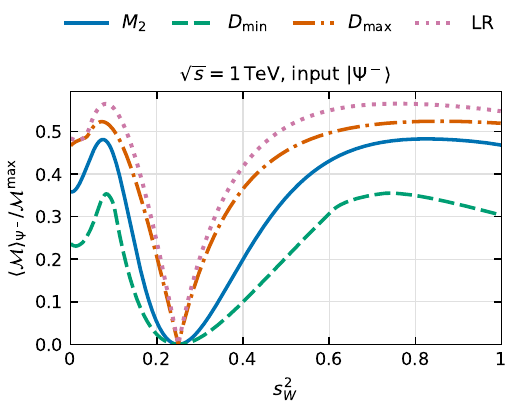}
  \caption{Average magic $\langle\mathcal{M}\rangle_{\Psi^-}$ of the
  outgoing spin state for the incoming singlet $\ket{\Psi^-}$ alone in
  M{\o}ller scattering, as a function of $s_W^2$ at
  $\sqrt{s}=1\,\mathrm{TeV}$, normalized to the two-qubit maximum of each
  measure.
All four
  measures vanish at $s_W^2=1/4$, where the outgoing state is
  $\ket{\Psi^-}$. $\mathfrak{D}_{\max}$ and LR rise linearly from there,
  $M_2$ and $\mathfrak{D}_{\min}$ quadratically, as in
  Eq.~\eqref{eq:mollerexp}.}
  \label{fig:psiminus}
\end{figure}

We neglect the electron mass, so that, with $E=\sqrt{s}/2$ the electron
energy, the Mandelstam variables are $t=-2E^2(1-\cos\Theta)$ and
$u=-2E^2(1+\cos\Theta)$, and define the sums of the photon and $Z$
propagators,
\begin{equation}
\begin{gathered}
  \Delta_\gamma=\frac1t+\frac1u,\\
  \Delta_Z=\frac{1}{t-m_Z^2+im_Z\Gamma_Z}+\frac{1}{u-m_Z^2+im_Z\Gamma_Z}.
\end{gathered}
\end{equation}
For small $g_V$, the normalized outgoing state for the input
$\ket{\Psi^-}$ is
\begin{equation}
\begin{aligned}
  \ket{\psi_{\mathrm{out}}}={}&\Big(1-\frac{|\alpha_1|^2g_V^2}{2}\Big)\ket{\Psi^-}\\
  &+(\alpha_1g_V+\alpha_2g_V^2)\ket{\chi_\Theta}+O(g_V^3),
\end{aligned}
  \label{eq:psiout}
\end{equation}
where $\ket{\chi_\Theta}=\sin\Theta\ket{\Phi^-}+\cos\Theta\ket{\Psi^+}$,
$\alpha_1=4\Delta_Z/(3\Delta_\gamma+\Delta_Z)$, and
$\alpha_2=-16\Delta_\gamma \Delta_Z/(3\Delta_\gamma+\Delta_Z)^2$. With
$\tau(\Theta)=\sin\Theta+|\cos\Theta|$, $\alpha_R=|\mathrm{Re}\,\alpha_1|$,
and $\alpha_I=|\mathrm{Im}\,\alpha_1|$, the four measures are
\begin{equation}
\begin{aligned}
  M_2&=\frac{4|\alpha_1|^2}{\ln2}\,g_V^2+O(g_V^3),\\
  \mathfrak{D}_{\min}&=\frac{|\alpha_1|^2}{\ln2}\,g_V^2+O(g_V^3),\\
  \mathfrak{D}_{\max}&=\frac{2\tau\,(\sqrt2|\alpha_1|-\alpha_R)}{\ln2}\,|g_V|+O(g_V^2),\\
  \mathrm{LR}&=\frac{\tau\,(\alpha_R+\alpha_I)}{\ln2}\,|g_V|+O(g_V^2).
\end{aligned}
\label{eq:mollerexp}
\end{equation}
The contribution of the singlet input over the full range of
$s_W^2$ is shown in Fig.~\ref{fig:psiminus} for $\sqrt{s}=1\,\mathrm{TeV}$.

\bibliography{ref}

\end{document}